# Enhancing the magnetoelectric coupling of $Co_4Nb_2O_9$[100] by substitution of Mg for Co

Zhen Li[1], Yiming Cao[2], Yin Wang[1,*], Ya Yang[3], Maolin Xiang[1], Youshuang Yu[1], Baojuan Kang[1], Jincang Zhang[1], Shixun Cao[1,†]

[1]Department of Physics, International Center for Quantum and Molecular Structures and Materials Genome Institute, Shanghai University, Shanghai 200444, China

[2]Center for Magnetic Materials, Devices & Key Laboratory for Advanced Functional and Low Dimensional Materials of Yunnan Higher Education Institute, Qujing Normal University, Qujing 655011, China

[3]School of Physics and Electronic Engineering, Xinyang Normal University, Xinyang 464000, China

*Corresponding authors. Email:* [*]*yinwang@shu.edu.cn*, [†]*sxcao@shu.edu.cn*

ABSTRACT

We report experimental studies on enhancing the magnetoelectric (ME) coupling of $Co_4Nb_2O_9$ by substitution of non-magnetic metal Mg for Co. A series of single crystal $Co_{4-x}Mg_xNb_2O_9$ ($x$=0, 1, 2, 3) with a single-phase corundum-type structure are synthesized by the optical floating zone method, and the good quality and crystallographic orientations of the synthesized samples are confirmed by Laue spots and sharp XRD peaks. Although the Néel temperature ($T_N$) of the Mg substituted crystals are slightly decreased from 27 K for pure $Co_4Nb_2O_9$ to 19 K and 11 K for $Co_3MgNb_2O_9$ and $Co_2Mg_2Nb_2O_9$, respectively, the ME coupling is double enhanced by Mg substitution when $x$=1. For the magnetic field (electric field) control of electric polarization (magnetization), the ME coefficient $\alpha_{ME}$ of $Co_3MgNb_2O_9$ is measured to be 12.8 ps/m (13.7 ps/m). These results indicate that the Mg substituted $Co_{4-x}Mg_xNb_2O_9$ ($x$=1) could serve as a potential material candidate for applications in future logic spintronics and logic devices.

# 1 Introduction

Combining spin degree of freedom into the conventional charge-based devices has generated much interesting physics and also led to the paradigm shifts of next generation electronics [1,2]. In this regard, the magnetoelectric (ME) effect played an important role, because ME coupling is considered as one of the most direct and effective way to realize the manipulation of electric polarization $P$ (magnetization $M$) by magnetic field $H$ (electric field $E$) in electronic devices [3] according to the relationship as expressed by $P_i = \alpha_{ij}H_j$ ($M_i = \alpha_{ij}E_j$) [4,5]. During the past several decades, many theoretical and experimental efforts have been devoted to design and fabricate novel ME materials in order to generate a larger ME coupling [3,6-7]. Among a series of ME materials, the linear ME materials which have no spontaneous polarization, have been widely investigated [8-13]. The family of $M_4A_2O_9$ materials, where $M$ is divalent transition metals and $A$ is Nb or Ta, is just a kind of linear ME material with honeycomb magnets and recently been more intensively investigated after its synthesis in 1961 [14] and the confirmation of its ME effect in 1972[15].

Among a number of $M_4A_2O_9$ materials that exhibit ME coupling [16-19], the most representative one is $Co_4Nb_2O_9$ (CNO), which crystallizes in a trigonal crystal structure (space group $P\bar{3}c1$) with two formula units per cell. $Co^{2+}$ ions in CNO occupy two inequivalent positions, denoted as Co1 and Co2, as shown in Fig. 1 (a). After the confirmation of the ME effect in CNO by Fischer *et al.* [15], the successful measurement of reasonably large ME coefficient α$_{ME}$ from pioneering efforts has been consistently demonstrated: Fang *et al.* [20] demonstrated the cross-coupling between magnetic and electric orders in polycrystalline CNO, where both magnetic field induced electric polarization and electric field controlled magnetization were observed; Yin *et al.* [21]reported that the origin of the linear ME effect can be ascribed to critical spin fluctuation. and Khanh *et al.* [22] reported that the ME response originates the reduction of symmetry due to magnetic order, respectively, in the crystal phase of CNO; Cao *et al.* [23] observed a nonlinear ME effect in CNO single crystal and found that the linear part of the α$_{ME}$ value is around 8 ps/m;

Dhanasekhar *et al.* [24] have partially substituted Co by Fe in CNO, and obtained a linear $\alpha_{ME}$ value of 7.6 ps/m; very recently, the magnetic structure of CNO is revealed by neutron diffraction, and more underlying physics of this materials could be explained [25].

Considering that element substitution is a very effective method to change or improve the materials properties, and the only substitution experiment work on the ME effect of CNO is Fe replacing Co by Dhanasekhar *et al.* [24], it is very worth trying to do more experimental exploration and study if substitution can induce larger ME coupling in CNO. In this paper, we substitute the magnetic element Co in CNO by a non-magnetic element Mg with different Mg concentrations *x*, and investigate the magnetic and electric properties of the Mg substituted single crystals. We focus on both magnetic field induced electric polarization and electric field controlled magnetization, and successfully observe the enhancing of the ME coupling in $Mg^{2+}$ substituted single crystals around *a* direction. In particular, we obtain a stronger capability of electric field control of magnetization in Mg substituted CNO, which is expected to play more important role in future technological applications.

2 Experimental details

$Co_{4-x}Mg_xNb_2O_9$ (CMNO) polycrystalline feed and seed rods were synthesized by the conventional solid-state reaction technique. A whole family of single crystals $Co_{4-x}Mg_xNb_2O_9$ with different Mg concentrations (*x*=0, 1, 2, 3*)* were successfully grown by an optical floating zone method (Crystal System Inc., type FZ-T-1000-H-VI-P-SH). The compounds of feed and seed rods were prepared by the raw materials $Co_3O_4$ (99.9%), MgO (99.9%) and $Nb_2O_5$ (99.9%), with the proper cation stoichiometry which was calculated from the target compound. The temperature of the molten zone was controlled by adjusting the power of lamps. During the growth process, the molten zone was moved upwards at a rate of 5 mm/h, with the seed rod (lower shaft) and the feed rod (upper shaft) counter rotating at 30 rpm in flow air. All of their crystallographic orientations were determined by an x-ray

Laue photograph (Try-SE. Co., Ltd.) with tungsten target (with the X-ray beam of 0.5mm in diameter) and additionally confirmed by standard XRD. X-ray diffraction （18KW D/MAX-2550）at λ=1.5406 Å was used to determine the crystal structures using the powders made by grinding part of the single crystal samples. Measurements of the magnetization were conducted using a Physical Property Measurement System (PPMS-9, Quantum Design Inc.) with VSM (Vibrating Sample Magnetometer) option.

The dielectric performance was conducted by using LCR meter (Agilent E4980A). The plates with the large [100] and [001] plane of 2*2 mm$^2$ were sliced out from the single crystal rod, and silver was painted as the electrodes. The pyroelectric current was collected using an electrometer (Keithley 6517B) after poling the sample in an electric and magnetic field, Electric polarization was obtained by integrating pyroelectric current with respect to time plus a careful exclusion of other possible contributions, such as thermally stimulated depolarization current from the de-trapped charges. The sample was cooled down to 10 K while applying a poling electric field of 0.5 MV/m along [100] direction. In order to release any charges accumulated on the sample surfaces, the sample was short-circuited for a long-enough time (1 h). During the recording of pyroelectric current, the sample was heated slowly at a warming rate of 2 K/min. Note that the magnetic field was applied throughout the cooling and warming processes. For further investigating the response of magnetization to electric field, the variation of magnetization as a function of a periodic electric field was also measured.

## 3 Results and discussionsc

Fig. 1 (a) shows the crystal structure of $Co_4Nb_2O_9$, where $Co^{2+}$ ions occupy two inequivalent positions, denoted as Co1 and Co2. The $Co^{2+}$ ions are substituted by $Mg^{2+}$ ions in our experiments and we will investigate the physical properties of the Mg substituted crystals in the following. Fig.1 (b) and (c) display the x-ray Laue photographs of $Co_3MgNb_2O_9$ for the *a* and *c* axes which were taken at room temperature. Rietveld method [26] as implemented in the FullProf program [27] was used for the refinement of the powder XRD data of single crystal $Co_3MgNb_2O_9$, and the results are shown in Fig. 1 (d). Fig.1 (e) shows the experimental powder XRD patterns of the CMNO family with different Mg concentration (*x*=1, 2, 3) at room temperature [28]. The diffraction patterns can be assigned to the single-phase modification of corundum trigonal crystal structure with space group $P\bar{3}c1$, and no impurity phase was detected, which indicates that high homogeneity of the single crystals was synthesized. Both XRD and Laue photography confirmed that our cutting plane are precisely perpendicular to *a* and *c* axes. Further neutron diffraction results indicate that Mg atoms prefer to replace Co1 rather than Co2 [29], which will help understand some results of this work.

The temperature dependence of magnetization $M_a(T)$ under a small magnetic field *H*=1 kOe applied along *a* axis of CMNO single crystals was measured and results are shown in Fig. 2 (a). The peak temperature around 27 K of green line can be defined as the Néel temperature of $Co_4Nb_2O_9$, which matches well with the earlier report [30]. And the Néel temperature of $Co_3MgNb_2O_9$ and $Co_2Mg_2Nb_2O_9$ are around 19 K and 11 K, respectively. But for $CoMg_3Nb_2O_9$ sample, we cannot find the anti-ferromagnetic (AFM) transition from *a* axis with the temperature down to 3 K. Clearly, with the increasing of the Mg concentration *x*, the Néel temperature of CMNO is decreased and finally no AFM state can be obtained when $Mg^{2+}$ was fully substituted for $Co^{2+}$. The temperature dependence of magnetization along *c* axis under a small magnetic field *H*=1 kOe of CMNO single crystals were also measured and results are shown in Fig. 2 (b). No AFM transition was observed below 50 K along *c* direction, which suggests that the magnetic moments are located in the *ab* plane for the Mg substituted crystals, rather than along *c* direction.

Having confirmed the crystal quality synthesized in our experiments and understood the AFM transition of CMNO samples, we begin to measure the *H/E* controlled *P/M* and calculate the ME coefficient $\alpha_{ME}$. Usually, an anomaly in dielectric constant is accompanied by a sudden change of electric polarization. Prior to the electrical measurements, the sample was first cooled down from 70 to 10 K with an electric field of 0.5 MV/m and a magnetic field of 0, 10, 30, 50 and 80 kOe along the *a* axis. After the ME fields-cooling, the pyroelectric current was recorded with increasing temperature without removing the external magnetic field. The value of electric polarization is obtained by integration of pyroelectric current with respect to time. The measurement of electric polarization as a function of temperature for different Mg concentration (*x*=0, 1, 2) was performed in magnetic field and poling electric field which was applied along [100] axis, and the results are shown in Fig. 3 (a), (b) and (c). It is obvious that no pyroelectric current is observed without external magnetic field. When magnetic field is applied, a pyroelectric current occurred at the temperature around $T_N$. The inset of Fig. 3 (b) profiles the dielectric constant as a function of temperature for $Co_3MgNb_2O_9$ at various magnetic fields along *a* axis. Under zero magnetic field, the dielectric constant shows no anomaly. However, peaks arise at $T_N$ when the external magnetic field is applied. With increasing the magnetic field, the peak in dielectric constant versus temperature curve becomes higher, indicating a magnetodielectric effect in these samples. Note that the $\alpha_{ME}$ value could be obtained by fitting the P-H curve as shown in Fig. 3 (d) ($Co_3MgNb_2O_9$). Although stable manipulation of electric polarization by magnetic field in $Co_2Mg_2Nb_2O_9$ (*x*=2) around *a* axis is not clearly observed [Fig. 3 (c)], we obtained a much larger $\alpha_{ME}$ value upto 12.8 ps/m for $Co_3MgNb_2O_9$ (*x*=1) around *a* axis [Fig. 3 (b) and (d)], compared to that of the pure CNO of 8 ps/m at *H* = 80 kOe. Notably, the $\alpha_{ME}$ value of $Co_3MgNb_2O_9$ is also larger than that of the well know ME materials like $Cr_2O_3$ (4 ps/m) [31], $MnTiO_3$ (2.6 ps/m) [32], $NdCrTiO_5$ (0.51 ps/m) [33]. Clearly, these results demonstrate that moderate Mg substitution (*x*=1) drastically enhances the ME coupling of CNO. As measured in our experiments and considering that Mg atoms prefer to replace Co1, Mg substitution has slightly changed the lattice parameters of CNO crystal and affected the Co2-O-Co2 bond angle and the Co2-Co2 interactions that contribute to the Dzyaloshinkii-Moriya interactions in CMNO according to the spin-current model [25]. These will induce the lattice distortion and led to the increasing of the ME coupling and magnetization of Mg substituted CMNO.

Considering that $Co_3MgNb_2O_9$ crystal have much larger magnetic field control of electric polarization, in the following we mainly investigate the electric field control of magnetization of $Co_3MgNb_2O_9$ samples.

As mentioned in the introduction section, the electric field control of magnetization is another important factor to evaluate the performance of a ME material. Fig. 4 (a) shows the magnetization versus temperature curves measured in heating process under various external electric field -0.5 MV/m, 0 MV/m and -0.5 MV/m. The magnetization above $T_N$ keeps unchanged under external electric field due to the disappearance of the polarization. However, for temperatures below $T_N$, each electric field value gives a different magnetization value. It is clear that the magnetization value increases or decreases the same magnitude with respect to the magnetization at zero electric field when opposite electric fields are applied. It is interesting to point out that below $T_N$ the magnetization can be tuned by a small electric field even under a weak cooling magnetic field. To approximately estimate the ME coefficients from the M-E curve, the fitting is conducted as $\mu_0 M = \mu_0 M(0) + \alpha_{ME} E$, the obtain results (see inset Fig.4 (a)) show that the $\alpha_{ME}$ equal to 13.7 ps/m, which is much larger than the pure CNO (8.27 ps/m) as in the previous report [17]. Slight difference between the fitted $\alpha_{ME}$ values for electric filed control of magnetization (12.8 ps/m at 10 K) and magnetic field control of electric polarization (13.7 ps/m at 15 K) may stem from the difference of the experimental conditions in our experiments.

To further confirm the performance of electric field control of magnetization for $Co_3MgNb_2O_9$, we measured the time dependence of magnetization under a square wave electric field of 0.5 MV/m, as shown in Fig. 4 (b). It is obvious that magnetization has a very quick and stable response with respect to external electric filed, indicating a promising application of this material for logic spintronics and electronics [34].

## 4 Conclusions

In conclusion, we have synthesized a series of single crystal $Co_{4-x}Mg_xNb_2O_9$ ($x$=0, 1, 2, 3) with a single-phase corundum-type structure by the optical floating zone method. Clear Laue spots and sharp XRD peaks confirm the good quality and crystallographic orientations of the synthesized samples. Although the Néel temperature ($T_N$) of the Mg substituted crystals are slightly decreased from 27 K for pure CNO to 19 K and 11 K for $Co_3MgNb_2O_9$ and $Co_2Mg_2Nb_2O_9$, respectively, the ME coupling is drastically enhanced by Mg substitution when $x$=1 in $Co_{4-x}Mg_xNb_2O_9$. For the magnetic field (electric field) control of electric polarization (magnetization), the ME coefficient $\alpha_{ME}$ of $Co_3MgNb_2O_9$ is doubled, compared to that of the pure CNO. More interestingly and importantly, the electric field control of magnetization of $Co_3MgNb_2O_9$ is enhanced. These results indicate that the Mg substituted $Co_{4-x}Mg_xNb_2O_9$ ($x$=1) could serve as a potential material candidate for applications in future logic spintronics and logic devices.

**Acknowledgements** This work is supported by the National Natural Science Foundation of China (Grant No. 11774217) and the Project for Applied Basic Research Programs of Yunnan Province (No. 2017FD142).

**Figure captions**

Fig.1 (a) Atomic structure of $Co_4Nb_2O_9$, $Co^{2+}$ ions occupy two inequivalent positions, denoted as Co1 and Co2. X-ray Laue photographs for (b) *a* axis and (c) *c* axis at room temperature for $Co_3MgNb_2O_9$, (d) The room temperature x-ray powder Rietveld refinement for $Co_3MgNb_2O_9$ and (e) X-ray diffraction patterns of the $C_{4-x}Mg_xNb_2O_9$ (*x*=1,2,3) single crystal family, the standard diffraction of $Co_4Nb_2O_9$ (PDF#38-1457) is plotted at the bottom panel.

Fig.2 (a) Magnetizations versus temperature curve for $Co_{4-x}Mg_xNb_2O_9$ (*x*=0, 1, 2, 3) along (a) *a* axis and (b) *c* axis under an applied magnetic field of 1 kOe in the temperature range from 3 to 50 K.

Fig.3 The temperature dependent polarization for (a) $Co_4Nb_2O_9$, (b) $Co_3MgNb_2O_9$ and (c) $Co_2Mg_2Nb_2O_9$ samples along *a* axis under different magnetic field. (d) The fitting of P-H curve for $Co_3MgNb_2O_9$ at 10 K. The inset in (b) shows the dielectric constant as a function of temperature at different *H* along *a* axis.

Fig.4 (a) The temperature dependence of magnetization under -0.5, 0, and 0.5 MV/m after ME cooling at 100 Oe. Inset shows the fitting of M-E curve for $Co_3MgNb_2O_9$ at 15 K. (b) The magnetization as a function of electric field after ME cooling at 100 Oe.

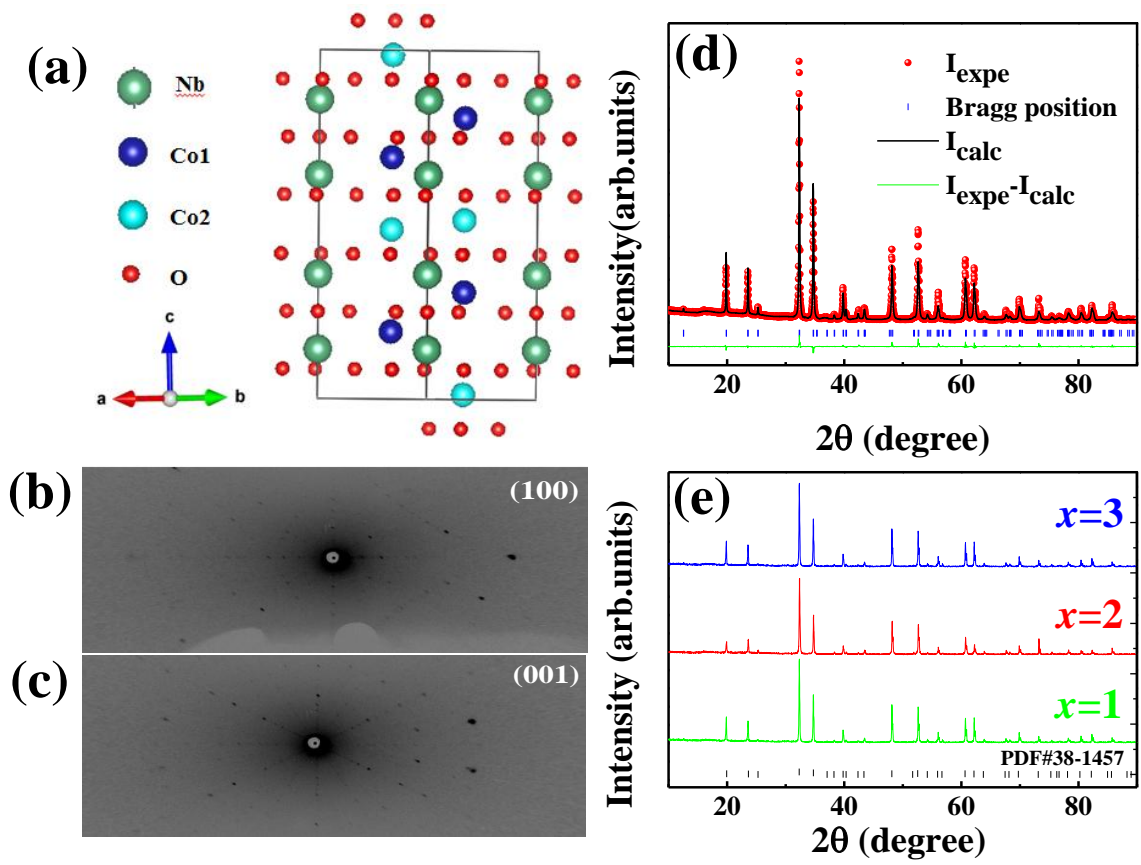

Figure 1

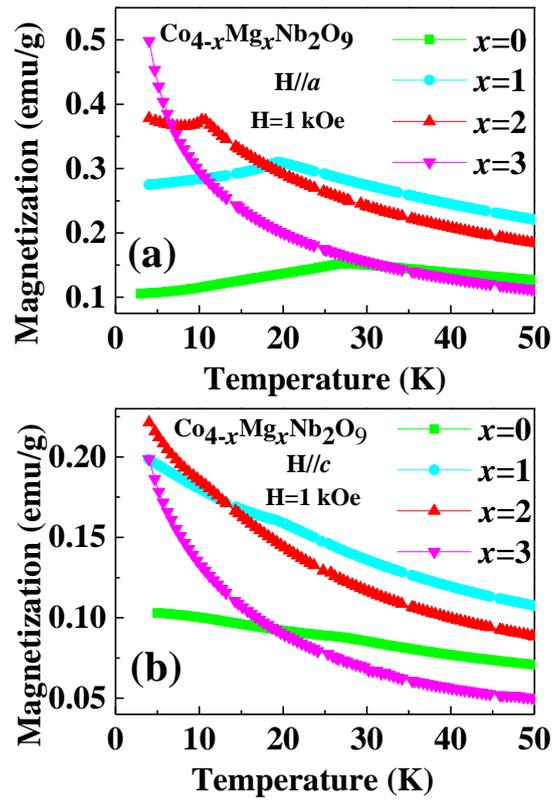

Figure 2

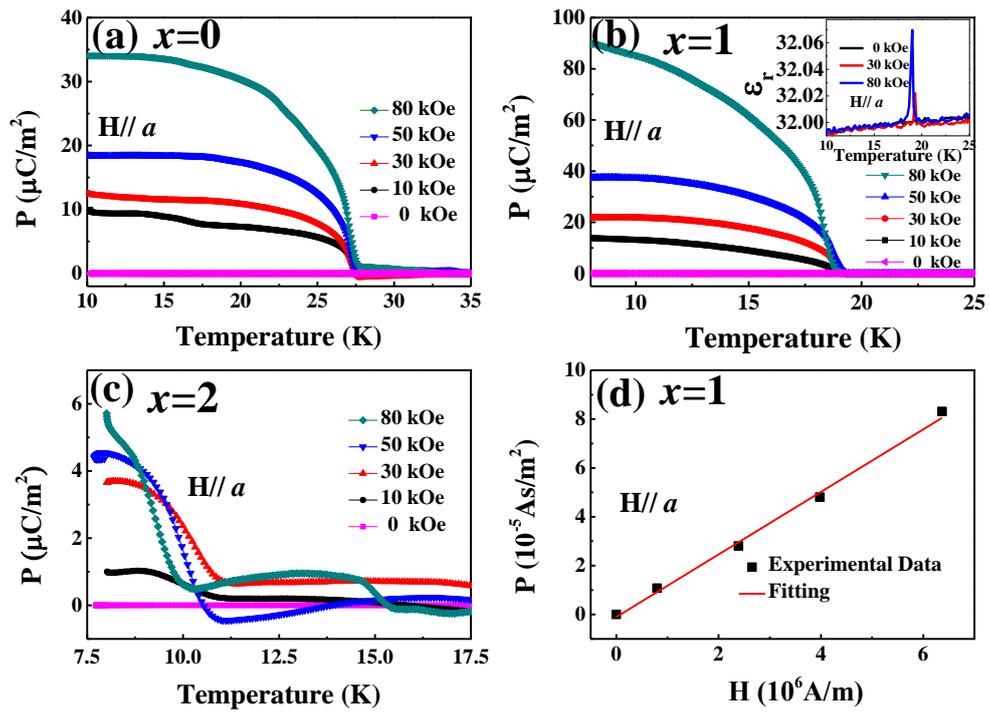

Figure 3

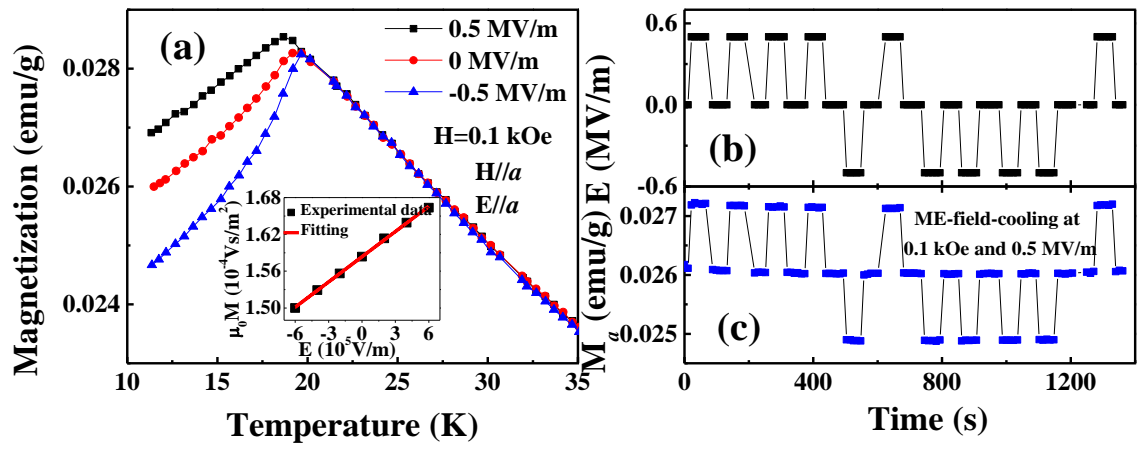

Figure 4